# Pressure-induced charge orders and their coupling to magnetism in hexagonal multiferroic LuFe$_2$O$_4$

Fengliang Liu[1,2,3#], Yiqing Hao[1#], Jinyang Ni[1#], Yongsheng Zhao[2], Dongzhou Zhang[4], Gilberto Fabbris[5], Daniel Haskel[5], Shaobo Cheng[6], Xiaoshan Xu[7], Lifeng Yin[1,8,9,10,11], Hongjun Xiang[1,9], Jun Zhao[1,9,10], Xujie Lü[2], Wenbin Wang[8,9,10,11,*], Jian Shen[1,8,9,10,11,*] and Wenge Yang[2,*]

[1]State Key Laboratory of Surface Physics and Department of Physics, Fudan University, Shanghai 200433, China

[2]Center for High Pressure Science and Technology Advanced Research (HPSTAR), Shanghai 201203, China

[3]Department of Physics, Nanchang University, Nanchang, 330031, China

[4]Hawaii Institute of Geophysics & Planetology, University of Hawaii Manoa, Honolulu, HI, USA.

[5]Advanced Photon Source, Argonne National Laboratory, Argonne, Illinois 60439, USA

[6]Department of Condensed Matter Physics and Materials Science, Brookhaven National Laboratory, Upton, NY 11973, USA

[7]Department of Physics and Astronomy, University of Nebraska, Lincoln, Nebraska 68588, USA

[8]Institute for Nanoelectronic Devices and Quantum Computing, Fudan University, Shanghai 200433, China

[9]Collaborative Innovation Center of Advanced Microstructures, Nanjing 210093, China

[10]Shanghai Qi Zhi Institute, Shanghai 200232, China

[11]Shanghai Research Center for Quantum Sciences, Shanghai 201315, China

[#]These authors contributed equally.

* Corresponding authors. **E-mail addresses:** wangwb@fudan.edu.cn (W.B.W.),

shenjian5494@fudan.edu.cn (J.S.), yangwg@hpstar.ac.cn (W.G.Y.).

Hexagonal LuFe$_2$O$_4$ is a promising charge-order (CO) driven multiferroic material with high charge and spin ordering temperatures. The coexisting charge and spin orders on Fe$^{3+}$/Fe$^{2+}$ sites result in novel magnetoelectric behaviors, but the coupling mechanism between the charge and spin orders remains elusive. Here, by tuning external pressure, we reveal three correlated spin-charge ordered phases in LuFe$_2$O$_4$: i) a centrosymmetric incommensurate three-dimensional CO with ferrimagnetism, ii) a non-centrosymmetric incommensurate quasi-two-dimensional CO with ferrimagnetism, and iii) a centrosymmetric commensurate CO with antiferromagnetism. Experimental in-situ single-crystal X-ray diffraction and X-ray magnetic circular dichroism measurements combined with density functional theory calculations suggest that the charge density redistribution caused by pressure-induced compression in the frustrated double-layer [$Fe_2O_4$] cluster is responsible for the correlated spin-charge phase transitions. The pressure-enhanced effective Coulomb interactions among Fe-Fe bonds drive the frustrated (1/3, 1/3) CO to a less frustrated (1/4, 1/4) CO, which induces the ferrimagnetic to antiferromagnetic transition. Our results not only elucidate the coupling mechanism among charge, spin and lattice degrees of freedom in $LuFe_2O_4$ but also provide a new way to tune the spin-charge orders in a highly controlled manner.

**INTRODUCTION**

Multiferroric materials have attracted lots of research interests during the last decades because of their great potential applications in electronic devices and spintronics[1,2]. Hexagonal $LuFe_2O_4$ is a promising candidate material for charge-order (CO) driven multiferroicity, which has been intensively investigated both from fundamental and applied perspectives[3-10]. Previous X-ray diffraction (XRD)[4] and transmission electron microscopy (TEM) measurements[11,12] have suggested that $LuFe_2O_4$ exhibits three-dimensional (3D) CO at ambient pressure, manifesting a periodic arrangement of low valence ($Fe^{2+}$) and high valence ($Fe^{3+}$) ions. Indication of the CO-driven ferroelectricity of $LuFe_2O_4$ was also revealed in the observation of the spontaneous electronic polarization above room temperature[3,13,14]. In addition to the 3D order of $Fe^{2+}$-$Fe^{3+}$ ions observed below the CO transition temperature $T_{CO}$ (~ 320 K), a quasi-2D (Q2D) ordering of $Fe^{2+}$-$Fe^{3+}$ was also observed above $T_{CO}$ persisting up to ~ 525 K[12]. Furthermore, neutron diffraction measurements have revealed ferrimagnetic order of Fe moments below the Neel transition temperature $T_N$ (~ 240 K)[15]. It is therefore suggested that the correlations between charge and magnetic order associated with $Fe^{2+}$ and $Fe^{3+}$ ions may play a crucial role in the multiferroicity in $LuFe_2O_4$[4,16-19].

Application of external pressure is a powerful and clean tool to gain deep insights into the interplay between the charge, magnetic and structural degrees of freedom, because the strong frustration involved in the spin and charge orders in the triangular lattice of $LuFe_2O_4$ may result in highly tunable ground states[16,19]. However, high pressure diffraction measurements are often difficult owing to the reduced beam flux and increased background caused by pressure cells. Previous neutron diffraction measurement on powder samples of $LuFe_2O_4$ revealed ~30% reduction of the ferrimagnetic ordered moments up to ~3 GPa[20], and X-ray powder diffraction showed indications for pressure-induced structural phase transitions[21]. However, the positions and intensities of superlattice reflections observed in powder diffraction showed non-systematic evolution with increasing pressure probably due poor powder averaging, thus preventing accurate description of the pressure-induced phases[21]. Moreover, neutron and X-ray powder diffraction measurements have revealed a high-pressure polymorph phase $LuFe_2O_4$-$hp$[22,23] above 12 GPa. The $LuFe_2O_4$-$hp$ phase retained its structure after pressure release, thus allowing ex-situ measurement at ambient pressure[22,23]. This $LuFe_2O_4$-$hp$ phase adopts a rectangular Fe lattice[23] and is not directly relevant to the frustrated triangular lattice of $LuFe_2O_4$ at ambient conditions. Therefore, despite the intensive efforts, the pressure dependent evolution of the frustrated charge and magnetic interactions in the multiferroic $LuFe_2O_4$ remains unclear.

Here, we investigated the evolution of charge and spin orders under pressure in $LuFe_2O_4$ using the in-situ high-pressure single-crystal X-ray diffraction (HP-SXD) and high-pressure X-ray magnetic circular dichroism (HP-XMCD) spectroscopy. A series of pressure-induced charge-order phases, including a Q2D CO phase below 5.5 GPa and a 3D CO phase at higher pressure (6.0 ~ 12.6 GPa), were identified by HP-SXD measurements. In addition, the HP-XMCD measurements suggest that the Q2D and 3D CO phases are associated with ferrimagnetic and antiferromagnetic order, respectively. The pressure induced charge- and magnetic-order phase transitions were further confirmed by density functional theory (DFT) calculations. These results suggest that the CO phases are intimately coupled with magnetism, both of which can be manipulated by external pressure in a highly controlled manner in hexagonal layered multiferroic $LuFe_2O_4$.

**RESULTS**

$LuFe_2O_4$ adopts a bi-layered triangular lattice structure with space group $R$-$3m$ (No.166) at ambient pressure. Considering the CO and the atomic displacement it induces, the trigonal lattice splits into three 120° twinned monoclinic lattices, each of which adopts the space group $C2/m$ (No.12). The transformation between the trigonal and monoclinic lattices is shown in Fig. 1a. The optimal stoichiometry of our sample was confirmed by magnetic susceptibility and transmission electron microscopes measurements at ambient pressure (Fig. S1). Our high-pressure synchrotron X-ray diffraction measurement of $LuFe_2O_4$ (Fig. S2) was carried out on a high quality single-crystalline sample. Within the representation of space group $C2/m$ (Fig. S3), the lattice parameters are $a = 5.957(2)$ Å, $b = 3.434(4)$ Å, $c = 8.642(3)$ Å, and $\beta = 103.28(2)°$ at 300 K and 0.8 GPa. The lattice parameters $a$, $b$, and $c$ shrink with increasing pressure. At 300 K and 12.6 GPa, the lattice parameters become $a = 5.809(5)$ Å, $b = 3.339(16)$ Å, $c = 8.374(11)$ Å, and $\beta = 103.39(9)°$ with a volume reduction of 8.2(6)%.

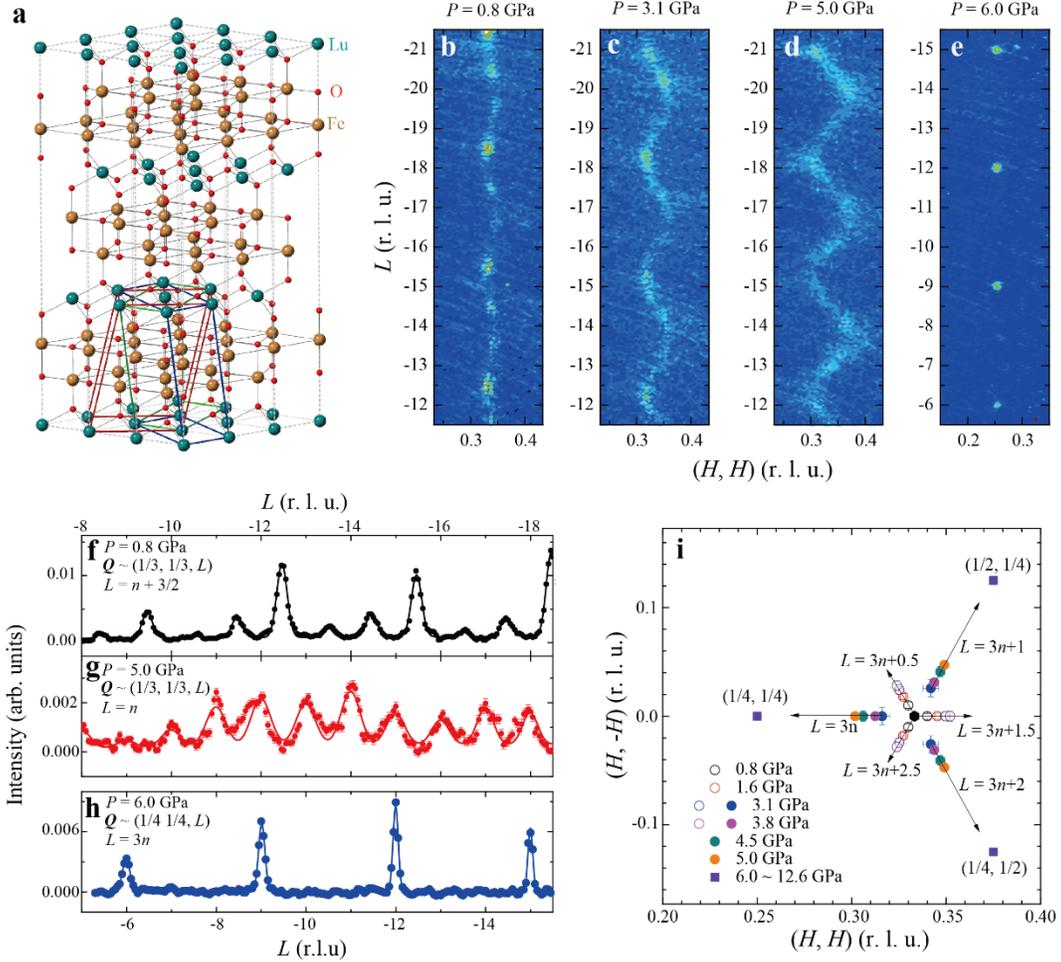

**Fig. 1 Crystal structure and charge order evolution of $LuFe_2O_4$ under high pressure.** (a) Crystal structure of $LuFe_2O_4$ at ambient pressure. Dotted lines represent the trigonal lattices. Red, blue and green solid lines represent three equivalent monoclinic lattices with different twinned directions. (b-e) Single crystal X-ray diffraction (SXD) intensities in the (*HHL*) scattering plane measured near the charge ordering (CO) superlattice peaks at room temperature at 0.8, 3.1, 5.0 and 6.0 GPa, respectively. (f-h) SXD intensity along the ***Q*** = (1/3, 1/3, *L*), (1/3, 1/3, *L*) and (1/4, 1/4, *L*) direction measured at 0.8 GPa, 5.0 GPa and 6.0 GPa, respectively. (I) Summary of the charge order peak positions in *HK*-plane measured from 0.8 to 12.6 GPa. The black hexagon in center indicates (*H*, *H*) = (1/3, 1/3).

Apart from the lattice shrinkage, a series of superlattice peaks were also observed with a wave vector $k_{AP}$ = (1/3, 1/3, 3/2)$_T$ in trigonal lattice or (0, 2/3, 1/2)$_M$ in monoclinic lattice at 0.8 GPa, as illustrated in Fig. 1b and Fig. 1f. This wave vector is close to that observed at ambient pressure where the charge order of low-valence ($Fe^{2+}$) and high-valence ($Fe^{3+}$) Fe ions forms a $\sqrt{3} \times \sqrt{3} \times 2$ super-lattice[4] (Fig. 2b). As pressure further increases, the superlattice peaks exhibit drastic broadening along the $L$-direction (Fig. 1(b-d)), indicating that the charge order becomes quasi-two-dimensional. Meanwhile, the maximum of peak intensity in L-direction moves from half integer ($L = n+1/2$) at ambient pressure to integer ($L = n$) at 5.0 GPa (Fig. 1(d, g) and Fig. S4), indicating that the inter-plane polarization emerges under pressure (Fig. 2d). Interestingly, similar quasi-two-dimensional charge order was also observed above the 3D charge ordering temperature of 320 K in $LuFe_2O_4$ at ambient pressure[4].

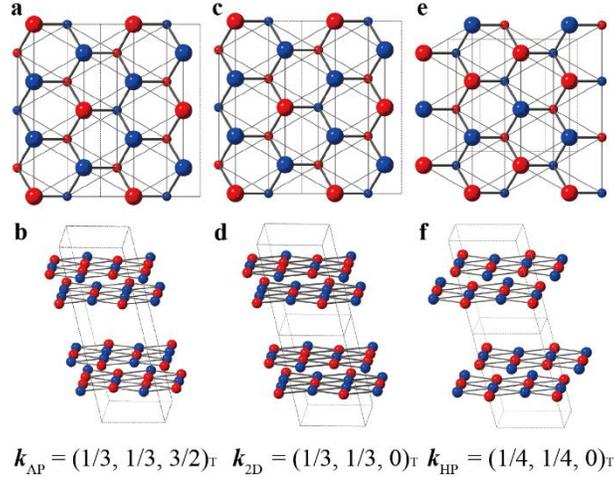

$k_{AP}$ = (1/3, 1/3, 3/2)$_T$   $k_{2D}$ = (1/3, 1/3, 0)$_T$   $k_{HP}$ = (1/4, 1/4, 0)$_T$

**Fig. 2 Charge order models of $LuFe_2O_4$ corresponding to observed SXD pattern from 0.8 GPa to 12.6 GPa.** Red and blue spheres represent $Fe^{2+}$ and $Fe^{3+}$ ions, respectively. Sphere size in (a, c, e) indicate the Fe ions in different planes. Thick and thin solid lines represent nearest neighbored (NN) and next nearest neighbored (NNN) Fe-Fe pairs, respectively. Dashed lines represent lattice boundaries. (a, b) Centrosymmetric CO-AP model with $k$ ~ (1/3, 1/3, 3/2). (c, d) Non-centrosymmetric CO-2D model with $k$ ~ (1/3, 1/3, 0) at $P$ = 5.0 GPa. Noted the inter-layer arrangements in (b) and (d) are different due to change in L-component of the wave vector. (e, f) CO-HP model for $k$ = (1/4, 1/4, 0), corresponding SXD patterns observed between 6.0 and 12.6 GPa.

A closer inspection on the diffraction pattern actually reveals a tiny incommensurability of the charge order wave vector [$k_{AP}$ = (1/3+δ, 1/3+δ, 3/2)$_T$ and $k_{2D}$ = (1/3-δ, 1/3-δ, 0)$_T$] at 0.8 GPa. The incommensurability increases dramatically with increasing pressure, and eventually reaches a commensurate position $k_{HP}$ = (1/4, 1/4, 0)$_T$ at pressures above 6.0 GPa. In contrast to the broadened diffraction patterns along the $L$-direction for the superstructure below 5.0 GPa (Fig. 1d), the (1/4, 1/4, 0)$_T$ phase shows sharp peak features both along the in-plane (HK) and out-of-plane ($L$) directions (Fig. 1(e, h)), indicating restoration of robust 3D order. This commensurate CO-HP phase persists up to 12.6 GPa without a major change in the primary structure.

The evolution of the incommensurability of the superlattice peaks in the HK-plane under various pressures is summarized in Fig. 1i. There are three 120-degree twinned charge orders in $LuFe_2O_4$, each with a unique $k$-vector, therefore the superlattice peak position differs for each charge order twin, forming a spiral-like diffraction pattern in reciprocal space. As pressure increases, the centers of charge order peaks move away from (1/3, 1/3) along the 120-degree directions in the HK-plane and eventually reach (1/4, 1/4), (1/2, 1/4), (1/4, 1/2) at 6.0 GPa. Therefore, the quasi-2D charge order phase observed in the 3.1 GPa to 5.0 GPa range can be regarded as an intermediate phase between the commensurate (1/3, 1/3) and (1/4, 1/4) charge-order phases.

We now discuss the possible charge order models for the high-pressure phases. We consider two valence states of Fe ions ($Fe^{2+}$ and $Fe^{3+}$). Based on diffraction data, we found that the charge order of $(1/4, 1/4, 0)_T$ phase can be best described with the $P_b2/c$ (BNS 13.71) black-white space group (Fig. 2(e,f) and Fig. S5). The CO phase models of CO-AP [$k_{AP} = (1/3, 1/3, 3/2)_T$], CO-2D [$k_{2D} = (1/3, 1/3, 0)_T$] and CO-HP [$k_{HP} = (1/4, 1/4, 0)_T$] are illustrated in Fig. 2(a,b), (c,d) and (e,f), respectively. These CO models are further supported by DFT calculation (Fig.4d). The evolution from $(1/3, 1/3)_T$ to $(1/4, 1/4)_T$ can be understood through the Coulomb interactions on different types of Fe-Fe bonds. In CO-AP and CO-2D phases, 5/9 of the nearest neighbor (NN), 2/3 of the 2nd NN and 5/9 of the 3rd NN Fe bonds are $Fe^{2+}$-$Fe^{3+}$ bonds. On the contrary, in CO-HP phase, 2/3 of the NN, 2/3 of the 2nd NN and 1/3 of the 3rd NN Fe bonds are $Fe^{2+}$-$Fe^{3+}$ bonds. Above observation provides a natural understanding, that the effective Coulomb interactions of NN and 3rd NN Fe-Fe pairs are tuned by the compression in frustrated triangular double-layer $Fe_2O_4$ structure due to increased pressure. We also calculated the interlayer next-nearest-neighbor ($V_{cNNN}$) and intralayer nearest-neighbor ($V_{abNN}$) Coulomb interactions at each pressure point. Our result shows that $V_{cNNN}/V_{abNN}$ remains almost unchanged in the region of 1 to 5 GPa, and drops sharply above 6 GPa where the $(1/4, 1/4)_T$ CO phase is favored[24] (Fig. S6).

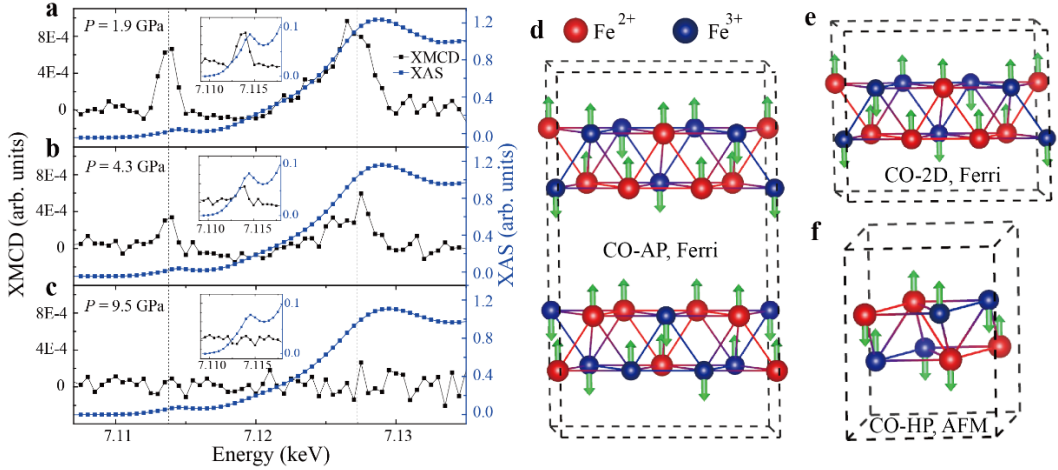

**Fig. 3 Magnetic properties of LuFe$_2$O$_4$ under pressure.** (a-c) Fe K-edge X-ray absorption near edge structure (XAS) and XMCD spectroscopy data measured at $T = 100$ K, $H = 5$ T and $P = 1.9$ GPa (a), 4.3 GPa (b), 9.5 GPa (c), respectively. The pressures correspond to the three observed charge order phases (CO-AP, CO-2D, CO-HP) in LuFe$_2$O$_4$. Insets of (a-c) show the XAS and XMCD data of pre-edge region at 1.9 GPa, 4.3 GPa and 9.5 GPa, respectively. (d) Ferrimagnetic model for the centrosymmetric CO-AP phase (e) Ferrimagnetic model for the non-centrosymmetric CO-2D phase (f) Antiferromagnetic model for the centrosymmetric CO-HP phase for DFT calculations in pressurized LuFe$_2$O$_4$. (Ferri: ferrimagnetic. AFM, anti-ferromagnetic).

In order to inform on the magnetic ground states associated with these charge-ordered phases, we utilized HP-XMCD spectroscopy to monitor the evolution of net magnetization of LuFe$_2$O$_4$ under pressure. Fig. 3a shows Fe K-edge isotropic absorption (XAS) and dichroic (XMCD) signals in LuFe$_2$O$_4$ at 100 K and 5 T. The measurements were performed at the Fe *K*-edge instead of the more commonly used Fe *L*-edge because soft X-ray MCD is incompatible with the highly absorbing diamond anvil cell environment[25-27]. The dichroic signal at 1.9 GPa consists of a positive peak in the pre-edge region (7114.0 eV) and another positive peak near the main rising edge peak (7129.0 eV) with larger intensity (Fig. 3a). With increasing pressure, these two dichroic peaks become much weaker and vanish at 9.5 GPa (Fig. 3c). This result indicates non-zero net magnetization at 1.9 GPa and 4.3 GPa, and zero net magnetization at 9.5 GPa. The peaks of Fe *K*-edge XAS and XMCD signals observed at 1.9 GPa are in agreement with those observed at ambient pressure (Fig. S7)[28]. Therefore, the magnetic order at 1.9 GPa is best interpreted as a ferrimagnetic structure which was unveiled by the previous L-edge XMCD[18] and neutron diffraction[15] measurements at ambient pressure. Meanwhile, the peak positions of dichroism peaks are unchanged at 1.9 GPa and 4.3 GPa (Fig. 3(a-c)). This result suggests that the ferrimagnetic arrangement of $Fe^{2+}$ and $Fe^{3+}$ moments is preserved at 4.3 GPa, but the net magnetization is gradually reduced by pressure. Insets of Fig. 3(a-c) illustrate the pre-edge region of XAS and XMCD signals. The positions of XAS pre-edge peak and leading edge do not shift by pressure, which indicates that the valence ratio of $Fe^{2+}$ and $Fe^{3+}$ remains unchanged (Fig. S8), in agreement with the CO models illustrated in Fig. 2.

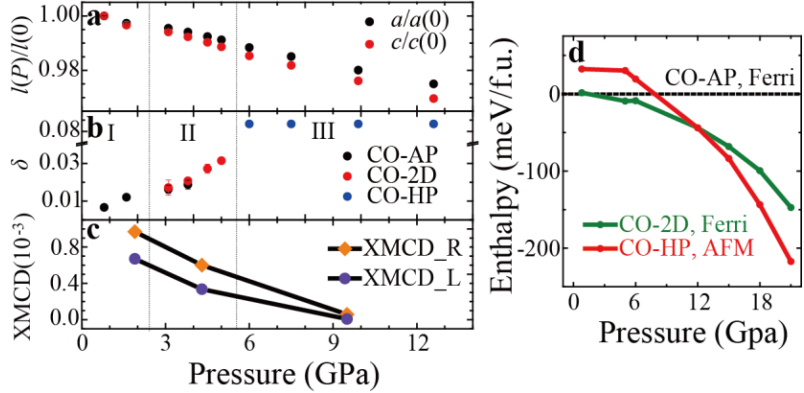

**Fig. 4 Phase diagram of LuFe₂O₄ under pressure.** (a) Relative reduction of lattice parameters *a* and *c* by increased pressure measured between 0.8 and 12.6 GPa. (b) Evolution of in-plane incommensurate CO wave vector $\delta$ by pressure. $\delta$ is defined by $(1/3+\delta, 1/3+\delta)$ in CO-AP and $(1/3-\delta, 1/3-\delta)$ in CO-2D. The commensurate CO wave vector $(1/4, 1/4)$ of CO-HP is represented by $\delta=1/12$. (c) The pressure dependence of XMCD spectral weight. (d) The pressure dependence of relative enthalpies between the charge order phases mentioned in Fig. 2 and Fig. 3d-f. (ferri: ferrimagnetic. AFM:anti-ferromagnetic).

More insight into the nature of the magnetic ground states associated with the CO phases in pressurized LuFe₂O₄ can be obtained by DFT calculations. For CO-AP phase and CO-2D phase, DFT calculations show that LuFe₂O₄ adopts the "2:1 ferrimagnetic state" as ground state. In this state, as shown in Fig 3(d, e), majority spin orientation consists of all $Fe^{2+}$ ions plus one-third of total $Fe^{3+}$ ions while the minority spin orientation consists of remaining $Fe^{3+}$ ions. For CO-HP phase, DFT calculation shows that the antiferromagnetic structure (Fig. 3f) is the magnetic ground state (Fig. S9). In this state, both $Fe^{2+}$ and $Fe^{3+}$ are evenly separated in opposite spin orientations, resulting in the centrosymmetric electromagnetic order. This result is consistent with the XMCD measurement, where the net magnetization is completely suppressed by pressure.

**DISCUSSIONS**

Combining the HP-SXD, HP-XMCD and DFT calculations, we have found a series of 3D-2D-3D charge order transitions in the hexagonal LuFe₂O₄. The emergence of the CO-HP phase at high pressure is accompanied by a ferri-magnetic to antiferromagnetic transition (Fig. 3(d-f)). We calculate the pressure dependence of the enthalpy for each spin-charge ordered phase using GGA+$U$ method[29,30], as illustrated in Fig. 4d. According to the calculation, at ambient pressure, the 3D centrosymmetric CO phase (CO-AP) combined with 2:1 ferrimagnetic order (Fig. 3d) is most favored. Then, as the lattice shrinks with applying hydrostatic pressure (Fig. 4a), a 2D non-centrosymmetric CO phase (CO-2D) with 2:1 ferrimagnetic order (Fig. 3e) becomes the ground state. At higher pressure, a 3D centro-symmetric CO phase (CO-HP) with antiferro-magnetic order (Fig. 3f) becomes the ground state. During the CO-AP to CO-2D phase transition, the Fe-Fe bonds within [Fe₂O₄] bilayer structure are compressed by pressure. Such compression allows charge transfer among low valence and high valence Fe sites in the frustrated lattice, which could result in damping of out-of-plane correlations and incommensurability in ab-plane (Fig. 1(b-d)). As the pressure further increases, the effective Coulomb interaction among Fe-Fe bonds reaches a critical point where the $Q$ = (1/3, 1/3) charge orders cannot be sustained. Thus the system evolves into a less frustrated $Q$ = (1/4, 1/4) CO-HP charge order where all Fe layers are charge neutral. Due to the strong coupling between charge and spin correlations, such drastic redistribution of charge density in turns causes substantial change in the magnitude of spin-spin interactions. As a result, the magnetic moments on $Fe^{2+}$ and $Fe^{3+}$ are re-arranged and the magnetic structure evolves from ferrimagnetism to anti-ferromagnetic order. The DFT calculations and experimental evidence indicate that the observed charge ordering and magnetic transitions are coupled and strongly correlated with the compression of the crystalline lattice. Our calculation reveals that the enthalpies of the three phases (CO-AP, CO-2D, CO-HP) are close to each other (Fig. 4d). This is not surprising as the double-layer triangular structure of $LuFe_2O_4$ induces strong frustration in both charge and spin exchange interactions.

In summary, we have conducted a systematic study on the evolution of the crystal structure, charge orders and spin orders in hexagonal $LuFe_2O_4$ under pressure using the in-situ HP-SXD and HP-XMCD measurements. With increasing pressure, the system exhibits three correlated charge ordered ground states: 1) the centrosymmetric 3D CO-AP phase with ferrimagnetic order, 2) the non-centrosymmetric CO-2D phase with ferrimagnetic order, and 3) the centrosymmetric 3D CO-HP phase with antiferromagnetic order. These results suggest strong coupling between charge and magnetic orders in hexagonal $LuFe_2O_4$. The evolution of the charge and magnetic order under pressure is further supported by the DFT calculations, which show that the phase transitions are the result of tuned frustrated charge and spin interactions induced by the compression of triangular $Fe_2O_4$ double layer structures. Our study proves that hydrostatic pressure is a powerful tool to unveil novel charge and magnetic states in hexagonal $LuFe_2O_4$ and other candidate multiferroics, and also provides new insights on realizing the potential charge-order induced multiferroicity in $LuFe_2O_4$.

**METHODS**
**Sample Preparation**
Single crystals of $LuFe_2O_4$ were grown by the floating-zone method using a $CO/CO_2$ (~ 2.7:1) mixed atmosphere to control oxygen stoichiometry. Our electron probe micro analyzer (EPMA) measurement on the single crystalline sample shows almost optimal stoichiometry of $Lu_{1.01(1)}Fe_2O_{3.97(4)}$.
**High-pressure measurement**
Single-crystal X-ray diffraction (SXD) experiment was conducted at synchrotron radiation beamline 13-BM-C, Advanced Photon Source (APS), Argonne National Lab (ANL), using monochromatic X-rays with 0.434 Å wavelength, with the pressure increasing from 0.8 GPa up to 14.5 GPa. The 1-deg step scan, wide-step scan and whole-range scan were performed with a scanning angle range of ±35 degrees at each pressure point[31]. Symmetry analysis was conducted using Sarah software[32] and Bilbao Crystallographic Server[33], and structure refinements were conducted using FULLPROF program suite[34]. High-pressure synchrotron X-ray Magnetic Circular Dichroism (XMCD) spectroscopy measurements with the energy scanning across the Fe K-edge were conducted at beamline 4-ID-D, APS, ANL. Circularly polarized X-rays were generated with a diamond phase retarder. To obtain the XMCD signal, the X-ray helicity was modulated at 13.1 Hz, and XMCD signals were detected with a phase lock-in amplifier. In addition, XMCD scans were repeated with opposite applied field direction to ensure artifact-free XMCD signals[35]. Corresponding to the three observed CO phases in pressurized $LuFe_2O_4$, the XMCD data were collected at 1.9 GPa, 4.3 GPa, and 9.5 GPa. In these measurements, magnetic field was set to +5T/-5T and temperature to 100 K. X-ray Absorption Spectroscopy (XAS) data is collected simultaneously during the XMCD measurement and obtained by averaging X-ray absorption for opposite X-ray helicities. The pressure was determined by the in-situ ruby fluorescence measurement system[36] with the pressure uncertainty less than 5%. An offline ruby system was used in the SXD measurement, before and after each manual pressure change at ambient temperature. For the low temperature XMCD measurements, online membrane and ruby fluorescence systems were used to apply and measure pressure, respectively (See details in supplementary Section S1).

**Density Functional Theory (DFT) calculations**

The first-principle density functional theory calculations are based on the projector augmented wave (PAW) method[37] encoded in the Vienna ab initio simulation package (VASP)[38]. The exchange-correlation functional of the Perdew-Becke-Erzenh (PBE)[39] form is adopted and the plane-wave cutoff energy is set to 500 eV. To properly describe the strong electron correlation in the Fe $3d$ states, the GGA plus on-site repulsion $U$ method (GGA+$U$)[29] was employed with the effective $U$ value ($U_{eff} = U - J$) of 4 eV. Calculations with various $U_{eff}$ show that the main results remain valid when $U_{eff}$ is varied between ~3.1 and 5.5 eV. The structural optimizations are carried out until the forces acting on atoms are smaller than 0.01 eVÅ$^{-1}$. To obtain the energy of each charge order structure, GGA+$U$ calculations were carried out in two steps[30]. For each charge order structure, we first optimize the chosen charge order structure using a large $U$ (e.g., $U_{eff}$ = 7.5 eV). Then we re-optimize the charge order structure with a smaller $U$ (i.e., $U_{eff}$ = 4 eV) using the converged charge densities obtained with the large $U$. It is noted that a larger $U_{eff}$ was employed only to generate an initial charge density with the desired charge order structure. The enthalpies of each charge order phase under various pressure are calculated by follow expression: $H = U – PV$, where $H$ is enthalpy, $U$ is total energy and $PV$ is product of pressure and volume. For each CO phase, atomic positions were relaxed while the lattice constants are constrained at the experimental values.

**DATA AVAILABILITY**

The data supporting of this study are available from the corresponding author upon reasonable request.

**ACKNOWLEDGMENTS**
This work was supported by National Natural Science Foundation of China (Grant No U1930401, 12074071, 51772184 and 11991060), and National Key Research Program of China (2016YFA0300702). We acknowledge Dr. Lili Zhang, Dr. Sheng Jiang and Dr. Aiguo Li of 15U1 at SSRF, and Dr. Changyong Park at HPCAT, APS, ANL for the technical supports on the high pressure XRD experiment. Gas loading by Sergey N. Tkachev is also acknowledged. HPCAT operations are supported by DOE-NNSA under Award No. DE-NA0001974 and DOE-BES under Award No. DE-FG02-99ER45775, with partial instrumentation funding by NSF. The SXD measurement was conducted at 13BM-C, APS, ANL. 13BM-C operation is supported by COMPRES through the Partnership for Extreme Crystallography (PX2) project, under NSF Cooperative Agreement EAR-1661511 and by GSECARS under NSF EAR-1634415. The XMCD experiment was conducted at beamline 4ID-D, APS, ANL. This research used resources of the Advanced Photon Source, a U.S. Department of Energy (DOE) Office of Science User Facility operated for the DOE Office of Science by Argonne National Laboratory under Contract No. DE-AC02-06CH11357.

**AUTHOR CONTRIBUTIONS**
Wenbin Wang, Jian Shen and Wenge Yang proposed and conceived this project. Fengliang Liu conducted the high-pressure XRD and XMCD experiments. Yiqing Hao, Fengliang Liu, Wenbin Wang, Jun Zhao, Lifeng Yin, and Wenge Yang analyzed the data. Jinyang Ni and Hongjun Xiang performed the DFT calculations. Dr. Dongzhou Zhang supported the SXD measurement, Drs. Gilberto Fabbris and Daniel Haskel supported the XMCD measurement, Dr. Yongsheng Zhao offered help in the XMCD measurements. Drs. Shaobo Cheng, Xiaoshan Xu and Xujie Lü offers valuable discussion during this project. Fengliang Liu, Yiqing Hao, Wenbin Wang, Wenge Yang and Jian Shen wrote the original version of the manuscript, and received feedbacks from all authors. Fengliang Liu, Yiqing Hao and Jinyang Ni contributed equally to this work.

**CONFLICT OF INTEREST**
The authors declare no conflict of interests.

**ADDITIONAL INFORMATION**
**Supplementary Information**
Supplementary data associated with this article can be found, in the supplementary files.